\documentclass[aps,twocolumn,showpacs,showkeys]{revtex4}
\usepackage[all]{xy}
\usepackage{graphicx, graphics} 
\usepackage{epsfig, subfigure}
\usepackage{latexsym, amssymb, amsmath, amsfonts}
\usepackage[english]{babel}
\usepackage[OT1]{fontenc}
\usepackage[latin1]{inputenc}
\usepackage{makeidx}

\begin{document}
\title[]{A geometrical viewpoint on the `Deformation Method'}
\author{V.~I.~Afonso$^{\,a}$\thanks{viafonso@df.ufcg.edu.br} and Diego Julio Cirilo-Lombardo$^{\,b,c}$}

\affiliation{$^{a}$Unidade Acadêmica de Física, UFCG, Campina Grande, Paraíba, Brazil. \\
$^{b}$ National Institute of Plasma Physics (INFIP),
Consejo Nacional de Investigaciones Científicas y Tecnicas (CONICET),
Facultad de Ciencias Exactas y Naturales, Universidad de Buenos Aires,
Cuidad Universitaria, Buenos Aires 1428, Argentina.\\
$^{c}$ Bogoliubov Laboratory of Theoretical Physics\\
Joint Institute for Nuclear Research, 141980 Dubna, Russia}

\begin{abstract}
At the classical level, redefinitions of the field content of a Lagrangian allow to rewrite an interacting model on a flat target space in the form of
a free field model (no potential term) on a curved target space. In the present work, we explicitly show that the idea of the `deformation method'
introduced in \cite{defmet}, can be interpreted in a very simple geometrical picture, in terms of a correspondence between the metrics of the curved target
spaces that arise in the free versions of the models. To that aim, we show explicit relations between the map function linking the fields and the metrics of the free models.
Also, the geometrical viewpoint puts in clear evidence the limitations of the method, restricted unavoidably to models of a single scalar field. 
However, by considering complex and even quaternionic field models, the applicability of the deformation procedure can be extended to systems 
with a content of two and four (constrained) real fields, respectively. To illustrate more possibilities, we also consider some supersymmetric models. 
In particular, a geometrical relation between the flat Minkowskian metric and a Fubini-Study space metric is explicitly constructed. 
This widening of the range of applicability of the method derives from the pure geometrical viewpoint here explored. 
\end{abstract}
\maketitle
{\bf 1. Introduction.} 

The `deformation method' --introduced in \cite{defmet} and explored in
several works \cite{otherdef}-- consist in a mapping between two classical scalar field models supporting {\it BPS} states, 
which links the corresponding solution spaces by a map (configuration space diffeomorphism) $f$ dubbed `deformation function'. 
This is possible whenever there exist a specific relation between the potentials governing the dynamics of the fields. 
That relation can be obtained by construction and, in that sense,
the method results much more efficient in finding new defect-like solutions than, for instance, 
the Rajaraman's `trial orbit method' of \cite{raja}.

As it is well known, the holonomy group of a semi-Riemannian manifold $(M,g)$ at a point $p \,(\in M)$ 
is defined as the group of parallel transports along loops based on $p$. 
It provides a powerful tool to study the geometric structure of the manifold and to discuss, 
for instance, the existence of parallel sections of geometric vector bundles. 
In his seminal paper \cite{berger}, Berger gave a list of
possible holonomy groups of simply connected (semi-)Riemannian manifolds
under the assumption that the group acts irreducibly on the tangent space at $p$. 
Precisely, such properties of these manifolds allow, at the classical level,
to map models comprising interacting fields on a flat target space, to
free fields models on a curved target space.
 
The main objective of the present work is to show that deformation method can be reinterpreted as a simple geometrical map
between the metrics of target spaces of the free versions of the models. 
The general idea of the mappings that can be constructed is synthesized in the diagram \eqref{diagram}. 
There it is shown, on the left side, `the original' (${\cal L}$) and the `deformed'  ($\tilde{\cal L}$) interacting models 
related by the deformation function $f$. The right side column shows the corresponding relation between the free versions
of the models (${\cal L}_g$ and $\tilde{\cal L}_g$ respectively), which are also connected by $f$, but now seen as a map 
between the target spaces metrics $g$ and $\tilde g$.

\vspace{-3mm}
\begin{equation}\begin{array}{ccl}
\nonumber{\mathcal{L}}\;\;[V(\phi )]\vspace{3pt} & \quad \longleftrightarrow
\quad & {\mathcal{L}}_{g}\;\;[(g(\theta )] \\ 
\vspace{3pt}\nonumber f\updownarrow \quad &  & \qquad \updownarrow f \\ 
\nonumber\tilde{\mathcal{L}}\;\;[\tilde{V}(\tilde{\phi})]\vspace{3pt} & 
\quad \longleftrightarrow \quad & \tilde{\mathcal{L}}_{\tilde g}\;\;[\tilde{g}(\tilde{\theta})]
\end{array}\label{diagram}
\end{equation}

In principle, the relation here proposed (as well as the original deformation method of \cite{defmet}) applies only to models of a single scalar field. 
However, considering complex --as in \cite{complexdef, kinknet}-- and quaternionic \cite{quat} field models, the procedure applicability can be enlarged
to comprise systems with a content of two and four (despite constrained) real field components, respectively. 

The results of the present work provide a deeper understanding of the deformation method and offer a mechanism to map geometries of interest 
in a wide range of contexts like Sugra, D-branes, Braneworlds, \emph{etc}.

As an illustration of the possibilities of application of the method, 
after reviewing the deformation procedure in a general complex case in Section II we apply, in Section III, the new geometrical interpretation to a `Quaternionic Wess-Zumino model',
developed in a previous work \cite{quat}. 
Then, in Section IV, we analyze the possibilities of bypassing the severe limitations of the method in its standard form by exploring the purely geometric viewpoint (no defect-type solutions involved). 
In particular, we consider supersymmetric models and work out the very important case of the Fubini-Study space. The summary and discussion of our results is left to Section VI.

\smallskip\textbf{2. Deformation procedure.} 

The dynamics of an interacting complex scalar field $\varphi$ on a bi-dimensional Minkowski spacetime is described by a Lagrangian of the form 
\begin{equation}  \label{complexLV}
\mathcal{L}= \tfrac{1}{2}\partial_\mu{\bar{\varphi}}\,\partial^\mu{\varphi}-V(\varphi,\bar{\varphi})\,,
\end{equation}
where the bar stands for complex conjugation and $V(\varphi,\bar{\varphi})$ is a scalar potential. 
As we are mainly concerned with the study and generation of defect-like solutions, 
we will consider here potentials showing spontaneous symmetry breaking.

In order to obtain physical (finite energy) solutions, the asymptotic  conditions: 
$|\varphi|\rightarrow v$ and ${|\dot\varphi|}\rightarrow 0$ with $V^\prime(v)=0$,
must be achieved as $x\rightarrow\pm\infty$ 
(the dot stands for derivation with respect to the spatial coordinate, $\dot\varphi=d\varphi/dx$, \emph{etc.}).
That is, solutions connect points of the vacua manifold (the set of minima of $V$).
Also, in the cases of interest (bosonic sector of supersymmetric theories), 
the potential $V$ can be put in terms of a `superpotential' $W$ 
(with $V(|\varphi|)=\tfrac12 \overline{W^{\prime }(\varphi)}\,{W^{\prime }(\varphi)}$). 
Under such conditions, the static \footnote{Non-static defect-like solutions for scalar field models 
are obtained by simply applying a Lorentz boost on the static ones.} 
second order variational field equation, $\ddot{\bar{\varphi}}-2\partial_{\varphi} V =0$, 
can be integrated once to obtain the first order equations
\begin{equation}  \label{eq:12}
\dot\varphi=\,e^{i \alpha}\,\overline{W^\prime(\varphi)},\qquad 
\dot{\bar\varphi}=\,e^{-i\alpha}\,W^\prime(\varphi).
\end{equation}
Here, $e^{i\alpha}$ corresponds to the $U(1)$ degeneracy on the choosing of the superpotential,
and $V_{0}=V(v)$, the `classical vacuum value' (value of the degenerate minima of the scalar potential) was set to zero for simplicity.

\medskip{\bf 2.1. `Standard' deformation.} The deformation procedure for complex field
models was explored in Ref. \cite{complexdef}. The deformed Lagrangian density has the general form 
$\tilde{\mathcal{L}}= (1/2) \partial^\mu\overline{\tilde\varphi}
\partial_\mu\tilde\varphi -\tilde V(\tilde\varphi,\overline{\tilde\varphi})$,
and describes the dynamics of the new complex field 
$\tilde\varphi=\tilde\varphi_{1}+i\tilde\varphi_{2}$, 
related to the original field $\varphi$ by a (at least) meromorphic function 
$f$, such that $\varphi=f(\tilde\varphi)=f_{1}(\tilde\varphi_{1},\tilde\varphi_{2})+if_{2}(\tilde\varphi_{1},\tilde\varphi_{2})$,
which, naturally, must fulfill Cauchy-Riemann conditions,
$\partial_{\tilde\varphi_{1}} f_{1} = \partial_{\tilde\varphi_{2}} f_{2}$ and $\partial_{\tilde\varphi_{2}}f_{1}=- \partial_{\tilde\varphi_{1}}f_{2}$.

The dynamics governed by $\mathcal{L}$ and $\tilde{\mathcal{L}}$ are
different, but we can relate $\tilde{V}(\tilde\varphi,\bar{\tilde\varphi})$
and $\tilde{W}(\tilde\varphi)$ to the original model by --see \cite{complexdef}--
\begin{equation}  \label{complexdefV}
\tilde{V}(\tilde\varphi,\bar{\tilde\varphi})=\frac{V(f(\tilde\varphi),
\overline{f(\tilde\varphi)})}{\left\vert f^{\prime}(\tilde\varphi)\right\vert ^{2}}
=\frac{1}{2}\tilde {W}^{\,\prime}(\tilde\varphi)\,
\overline{\tilde{W}^{\,\prime}(\tilde\varphi)}
\end{equation}
The corresponding first-order equations for the new field $\tilde\varphi$ read exaclty as \eqref{eq:12},
 but now with the superpotential $\tilde W$ determined by \eqref{complexdefV}.

The $BPS$ kink solutions for this system are obtained from the solutions of the undeformed first order equations,
 by simply taking the inverse map \cite{complexdef}, that is,  
\begin{equation}
\tilde\varphi^{K}(x)=f^{-1}(\varphi^{K}(x)).
\end{equation}

{\bf 2.2. Free model.} The free action corresponding to the model 
\eqref{complexLV} requires the introduction of a Kähler manifold with
metric $g_{z\bar{z}}$ 
\begin{equation}
\mathcal{I}_{g}=\frac{1}{2}\int dx^{n}g_{z\bar{z}}
\,\dot{\theta^{z}}\dot{\bar{\theta}}^{\bar z} ,
\end{equation}
where $\dot\theta^z=d\theta^z/dx$. Variation of this action leads to the field equations 
\begin{equation}  \label{eq:eom}
\ddot\theta^{w}+ g^{w\bar z}\left[ g_{z\bar z,w} \dot\theta ^{w}+g_{z\bar z,\bar w} 
\dot{\bar\theta}^{\bar w}-g_{z\bar w,\bar z}\dot{\bar\theta}^{\bar w}\right] \dot\theta^{z}=0
\end{equation}

Being $g_{zw}$ the metric of a Khäler manifold, the last two terms in \eqref{eq:eom} cancel, 
as in complex coordinates $g_{z\bar z} = g_{\bar z z}$ and $g_{zz} = g_{\bar z\bar z} = 0$. 
Therefore, we obtain 
\begin{equation}\label{relat}
g_{z\bar z} \ddot\theta^{z}+ g_{z\bar z,w} \dot\theta^{w} \dot\theta^{z}=0
\end{equation}
We can also drop the indices and put the expression in the simpler form 
$\ddot{\theta}+\partial_{\theta}\ln{g}\,\dot{\theta}^{2}=0$, 
which can be integrated to obtain the relation 
\begin{equation}  \label{eq:dos}
\dot\theta\dot{\bar\theta} = \mathcal{C}_{0}\, g^{-1}(\theta,\bar\theta),
\end{equation}
with $\mathcal{C}_{0} \in \mathbb{C}$ an integration constant.

If the interacting and the free models are to describe the same physics, 
their field equations should match after making explicit the relation between the fields $\phi$ and $\theta$. 
Now, putting \eqref{eq:12} in the form $\dot\varphi \dot{\tilde\varphi} = V(\varphi,\varphi)-V_{0}\,$,
and comparing with \eqref{eq:dos}, we can establish the relations between the curved (free) and the flat (interacting) versions
\begin{eqnarray} \label{eq:freecomplex} 
\dot\theta^2_1+\dot\theta^2_2&=&\dot\varphi^2_1+\dot\varphi^2_2\\
\mathcal{C}_{0}g^{-1}(\theta,\bar\theta)&=&V(\varphi,\bar{\varphi})-V_{0},
\label{eq:freecomplexb}
\end{eqnarray}
where we have made explicit the 
real and imaginary components of the fields. 
The simplest solution arises by taking 
$\theta_1=\varphi_1+\mathcal{C}_1$ and $\theta_2=\varphi_2+\mathcal{C}_2$ or, simply,
\begin{equation} \label{trivialsol}
\theta=\varphi+ \mathcal{C} 
\end{equation}
with $\mathcal{C}\in \mathbb{C}$, an arbitrary constant.

\smallskip{\bf 2.3. `Geometric map' deformation.} Now, starting with the free
version of the complex deformed model ($\tilde{\mathcal{L}}_{g}~=~\frac{1}{2} 
\tilde{g}_{z\bar z}\,\dot{\tilde{\theta}} ^{z}\dot{\bar{\tilde{\theta}}}^{\bar z}$),
from prescription \eqref{complexdefV}, and using relations analogous to %
\eqref{eq:freecomplex} and \eqref{eq:freecomplexb} for the deformed fields $%
\tilde\theta$ and $\tilde\varphi$, we can write 
\begin{equation}
\tilde C_{0}\tilde g^{-1}(\tilde\theta,\bar{\tilde\theta}) = 
\frac{V(f(\tilde\varphi_\pm),\overline{f(\tilde\varphi_\pm)})}{\left\vert
f^{\prime}(\tilde \varphi_\pm)\right\vert ^{2}} -\frac{V_{0}}{\left\vert
f^{\prime}(\tilde \varphi_{\pm})\right\vert ^{2}}
\end{equation}
or, using \eqref{eq:freecomplex}, 
\begin{equation}  \label{eq:freecomplexdef}
\tilde C_{0}\,\tilde g^{-1}(\tilde\theta,\bar{\tilde\theta}) = \frac{C_{0}\,
g^{-1}(\theta,\bar{\theta})+V_{0}}{\left\vert
f^{\prime}(\tilde\varphi)\right\vert ^{2}} -\frac{V_{0}}{\left\vert
f^{\prime}(\tilde\varphi_{\pm})\right\vert ^{2}}.
\end{equation}
Note that, depending on the values of the potential at de minima, 
these expressions can annihilate giving rise to singular points in the target space,
as we will see later.

Finally, using relation \eqref{trivialsol} and letting constants aside for simplicity, we
obtain an ODE for the function $f$ 
\begin{equation}  \label{eq:fcplx}
f^{\prime}(\tilde\varphi) \overline{f^{\prime}(\tilde\varphi)}=
|f^{\,\prime}(\tilde{\varphi})|^{2}= 
{\tilde g(\tilde\theta,\bar{\tilde\theta})}/{g(\theta,\bar{\theta})},
\end{equation}
which implements the map between the metrics of the free complex models.

\smallskip{\bf 3. Geometrical Interpretation.} 

Sigma models can be generically described by an action of the form 
\footnote{In the case of the Born-Infeld theoretical framework, 
$S=\int \sqrt{G\left(\phi \right)}$, where the determinant of the metric 
of the target space is the wedge product of the Maurer-Cartan forms.} 
\begin{equation}  \label{sigmaG}
S=\int d\phi\, G_{ij}\left( \phi \right) \ast d\phi
\end{equation}
Classically and geometrically speaking, we can think $G_{ij}$ as a metric on
the target space ($T$). In such direction, relation \eqref{eq:fcplx}
can be understood as a differential equation determining the simplest map $f$
connecting two given metrics $g$ and $\tilde{g}$, describing the target
spaces geometries of two distinct free field models. That is, 
\begin{equation}  \label{fgg}
\int df {\left[ g(f)\right]^{1/2}}=\pm \int d\tilde{\phi} \,[ \tilde{g}(\tilde{\phi})]^{1/2}.
\end{equation}
This expression reminds us the invariance of the action $S$ under Diff$(T): 
\phi \rightarrow \phi^{\prime}\left(\phi\right) \,\Rightarrow\,
G_{ij}\left(\phi\right)\rightarrow G_{i^{\prime}j^{\prime}}\left(\phi^{\prime}\right)$.
This is reminiscent of the determinantal character of the Sigma model as
measure, and expresses that a Sigma model is defined by an equivalence class
of metrics.

The other important ingredient is the invariance under reparameterization
implied by \eqref{fgg}, due to the square root (\emph{e.g.} Nambu-Goto/Barbashov-Chernikov action).

Taking a common canonical basis $\chi $ in the target space (where the
functional quantities have explicit dependence) we have 
$\tilde{g}(\chi )=g(f(\chi ))[f^{\,\prime }]^{2}$  and 
$\tilde{g}(\chi )=g(\tilde{\phi}\left( \chi \right) )[\tilde{\phi}^{\prime}]^{2}$.
So, immediately, we obtain 
\begin{equation}\label{geomfprime}
\frac{g(\tilde{\phi})}{g(f)}=\frac{[f^{\,\prime }]^{2}}{[\tilde{\phi}^{\prime }]^{2}}
=\left[ \frac{df}{d\tilde{\phi}}\right] ^{2}
\end{equation}

In the original prescription of the method, the function $f$ connects the field solutions (and potentials)
of two different models ($\phi=f(\tilde{\phi}))$. 
In our new approach, $f$ connects different metrics related by the
equivalence class of diffeomorphisms in the target plus reparametrization invariance.
This puts in evidence the reason why the original procedure worked
in all the cases treated in several articles:
it can be achieved only when considering just a single scalar field. 
The freedom reduces then, to the choice of the field of numbers over which
this single quantity is defined (namely, reals, complex or quaternions).

Notice also that, from a perturbative quantum mechanical point of view, 
$G_{ij}(X)$ is sometimes thought as the sum of an infinite number of coupling
constants: $G_{ij}\left( X\right) =G_{ij}^{0}+G_{ij,k}^{1}X^{k}+...$ These
constants do not preserve the invariance mentioned above, leading to a
breaking of the full symmetry of the theory at the quantum (perturbative)
level.

\smallskip {\bf 3.1 Topological sectors and singularities.} Another important
point to remark here is that, as stressed before, the asymptotic values of
the field solutions lead to singular points of the target space metric, as 
$g^{-1}(\phi _{\pm }^{K}(x)\rightarrow v_{\pm })\rightarrow V(v_{\pm})-V_{0}=0$ 
and, naturally, a similar behavior should be expected of the `free deformed model's metric $\tilde{g}$
\footnote{Note, however, that the behavior of $f^{\prime }$ could perhaps affect the
divergence order of $\tilde{g}$, as can be seen from \eqref{eq:fcplx}.}.
Thus, while degenerated minima of the original potential are mapped to degenerated
minima of the deformed potential, in the free model case, 
they are all mapped to the same singular point of the metric.
This is so as they depend on the values of the potential calculated at the minima which, 
by assumption, are degenerated.

Summarizing, topological sectors of the field models have their counterpart as
disconnected regions on the target space, separated by singularities.
This is a direct consequence of the fact that we are dealing with single field models.
This fact may result relevant for some supergravity theories, which demand such a type of
constructions -- see, for instance, \cite{sugraDW}.

\smallskip{\bf 3.2. Toy example: Sine-Gordon as $\protect\phi^4$ model deformation.}
Before attacking the quaternionic case, let us illustrate how the mechanism works by analyzing a trivial example. 
Consider the two very well known real scalar field models $\phi^4$ and sine-Gordon. 
As shown in \cite{fi4sGdef}, these two models can be linked by the deformation method. 
This is accomplished by simply taking the $\phi^4$ model, 
$V(\phi)=\tfrac12(1-\phi^2)^2$, 
which supports two $\mathbb{Z}_2$ kink solutions $\phi^K_\pm(x)=\pm\tanh(x)$
and applying the deformation function 
$f(\tilde\phi)=\sin(\tilde\phi)$.
Using prescription \eqref{complexdefV}, we immediately arrive to 
\begin{equation}
{\mathcal{V}}(\tilde\phi)= \frac{V(f(\tilde\phi))}{[f^\prime(\tilde\phi)]^2}=%
\frac{\frac12(1- \sin(\tilde\phi)^2)^2}{\cos^2(\tilde\phi)}=\tfrac12
\cos^2(\tilde\phi),
\end{equation}
which is the sine-Gordon potential \footnote{This corresponding to the more general expression 
$V(\phi)=\alpha \cos(\beta\phi)+|\alpha|$, fixing the parameters to the values $\alpha=1/4$, $\beta=2$.}. 
The sine-Gordon solutions are then obtained by inverting the $f$ function, namely 
\begin{equation}  \label{solbasG}
\tilde\phi_k(x)=f_k^{-1}(\phi^K(x))=\pm\arcsin(\tanh(x))+ k \pi,
\end{equation}
where $k\!\in\! \mathbb{Z}$ specifies the branch \footnote{Again, this result is consistent with the general form o the sine-Gordon kink solutions,
$\phi_n(x)= 2\arctan(e^x)+(2n+1)\tfrac{\pi}{2}$, taking the specific values of $\alpha =1/4$, $\beta =2$, 
and $n= k-1$, as can be easily checked.} of the inverse of $f$.

Note that, while the kink solutions of the $\phi^4$ model connect the single
topological sector defined by the pair of degenerated minima of the
potential ($v_+=+1$ and $v_-=-1$), the sine-Gordon model presents an
infinite number of topological solutions, one for each topological sector,
connecting adjacent minima ($(k-1/2)\pi$ and $(k+1/2)\pi$) -- see Fig. 1 in 
\cite{fi4sGdef}.

\smallskip Let us now consider the free versions of the models. 
From relations \eqref{trivialsol} and \eqref{eq:freecomplexdef} we have, for the $\phi^4$ model 
\begin{equation}
g_{\lambda}(\pm\phi+C_{1})=C_{0}(1-\phi^2)^{-2},
\end{equation}
and for the sine-Gordon model, 
\begin{equation}
\tilde{g}_{sG}(\pm\tilde\phi+\tilde C_{1})=\tilde C_{0}\cos^{-2}(\tilde\phi),
\end{equation}
with $\theta =\pm\phi+C_{1}$ and $\tilde\theta =\pm\tilde\phi+\tilde C_{1}$, respectively.

Now, from \eqref{fgg}, the deformation function connecting these free
models is given by 
\begin{eqnarray}
\int\!df\!\sqrt{g(f)}\!=\!\int\! \tfrac{df}{1-f^{2}} \!\!&\!\!=\!\!&\!\!\pm\!\int\! 
\tfrac{d\tilde{\phi}}{\cos (\tilde{\phi})}\!=
\!\pm\!\!\int\!\!d\tilde{\phi}\sqrt{\tilde{g}(\tilde{\phi})}\; 
\notag \\
\Rightarrow\quad f&=&\sin (\tilde{\phi})
\end{eqnarray}
That is, by using the `geometric picture' we obtain exactly the same mapping function that connects the
interacting version of the models given in \cite{fi4sGdef}.
Note also in the solution above the clear interplay between 
the minima of the potential and the singularities at the geometrical level or, better, between
the topological sectors and the causally disconnected regions of the target space determined by the singularities. 

\medskip{\bf 4. Quaternionic Field Model Deformation.} 

In a previous work \cite{quat}, we have shown genuine $BPS$ solutions for
a quaternionic Wess-Zumino model ($QWZ$), in the context of hyper-Kähler
structures. Even when this is a standard choice, we found valid to work with the $QWZ$ model
as it serves as the basic prototype for any analysis involving hypercomplex quaternionic structures, 
appearing in several areas of modern theoretical physics.

In general, domain walls are co-dimension one solutions, that separate the
spacetime into regions corresponding to different vacua. In the simplest
case, a domain wall is supported by a gauge potential that couples to its
world volume and the field strength of this gauge potential is dual to a
cosmological constant. In a more general setting, with nontrivial couplings
to scalar fields, this cosmological constant appears as an extremum of the
potential term in the Lagrangian. The resulting solution describes then a
flow towards an extremum and, if the potential possesses several extrema,
the solution may interpolate between them.
In the present section we will analyze a domain wall-like solution, analogous to the one
obtained in \cite{nitta}, developing a connection between solutions obtained
in our previous work \cite{quat}.

\smallskip{\bf 4.1. Hyper-Kähler domain wall solution. }
As starting point, consider a single quaternionic field governed by the
Generalized Quaternionic Lagrangian density ($GQL$) of the form -- see \cite{quat} -- 
\begin{equation}  \label{GQL}
\mathcal{L}=\tfrac{1}{2}\overline{\Pi q}\,\Pi q-\tfrac{1}{2}\left\vert
W^{\prime}(q) \right\vert ^{2},
\end{equation}
where the Cauchy-Fueter operator is given by $\Pi\equiv\widehat{i^{0}}%
\partial_{0}-\widehat{i^{1}}\partial_{1}-\widehat{i^{2}}\partial_{2}-%
\widehat{i^{3}}\partial_{3}$, with $\partial_{0}\equiv{\partial }/{\partial
x^{0}}$ and $\partial_{k}\equiv{\partial}/{\partial x^{k}}$, while $\widehat{%
i^{0}}=\mathbb{I}$ and $\widehat{i^{k}}$ ($k=1,2,3$) obey the standard
quaternionic algebra. Einstein's summation convention is adopted 
while the prime indicates derivative with respect to the argument of the function.

The quaternionic Wess-Zumino superpotential takes the form 
\begin{equation}  \label{WZpot}
W^{\prime }(q)=n-q^{N}=n-(q_{0}+\widehat{i^{k}}\,q_{k})^{N},
\end{equation}%
where $N\in\mathbb{Z}$. In general, $n\in \mathbb{H}$, 
but through this work we will take $n\in \mathbb{C}$ or in its subfields.

The corresponding vacuum (minima) manifold is described by the set of the 
$N$-roots of $n$ in the field of the quaternions, \emph{i.e.} $S^{2}$ spheres.
The first order equation, $\Pi\, q=\overline{W^{\prime}(q)}$, for our $QWZ$
potential \eqref{WZpot} reads 
\begin{equation}  \label{1stord}
\frac{dq}{dx} =n-\overline{q}^{N}=n-(q_{0}-\widehat{i^{k}}q_{k})^{N}
\end{equation}
This expression, with $x$ identified below, arises from the relation between
the left regular superpotential $W(q)$ and the $BPS$ conditions \cite{quat}.

\smallskip{\bfseries 4.2. Case $N=2$. New $BPS$ solution} (Non-commutative base space).  
We now present here a new $BPS$ quaternionic solution (not worked out in \cite{quat}), 
for the case $N=2$, with the fields on a quaternionic base manifold
(non-commutative spacetime equivalent) as target space. This is obtained by
identifying the spacetime spatial coordinate $x$ with one of the complex
directions of the quaternionic manifold. In this case we take $x\rightarrow 
\widehat{i}_{3}X^{3} $ (\emph{i.e.} $x\in SU(2)$) and, consequently, $\Pi=-%
\widehat{i^{3}}\partial_{3}$. As a consequence of this choice, the spacetime
assumes the structure $S_1\otimes O(3)\sim S_1 \otimes SU(2)$.

\begin{figure}[h!tb]\center
\includegraphics[{width=8cm}]{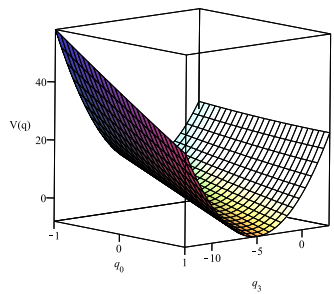} 
\caption{Quaternionic potential $V(q)$ ($n=1$).}
\label{fig:Vq}
\end{figure}

The potential for the $N=2$ case of our $QWZ$ model takes the form 
\begin{eqnarray}  \label{VN2}
V&=&\tfrac{1}{2}(n-\overline{q}^{2})(n-q^{2}) \\
&=&\tfrac{1}{2} n^2-n (q_0^2-q_3^2)+\tfrac{1}{2} (q_0^2+q_3^2)^2  \notag
\end{eqnarray}
Fig. \ref{fig:Vq} shows the form of potential \eqref{VN2} as a function of two of the
quaternionic components ($q_0$ and $q_3$). 

The first order equation \eqref{1stord} takes the form 
\begin{equation}  \label{1stordN2}
\frac{dq}{dX_3} =n-(q_{0}^{2}-q_{i}^{2}-2\,\widehat{i^{k}}q_{k}\,
q_{0}),\quad n\in\mathbb{Z}.
\end{equation}
Breaking equation \eqref{1stordN2} into its scalar and vectorial parts, we obtain
the system of equations 
\begin{equation}  \label{sysq}
\left\{%
\begin{array}{rclrcl}
\frac{dq_{0}}{dX_{3}} & = & -2q_{0}q_{3} \,, & \qquad \frac{dq_{1}}{dX_{3}}
& = & -2q_{0}q_{2} \\ \\
\frac{dq_{3}}{dX_{3}} & = & n-q_{0}^{2}+q_{i}^{2} \,, & \qquad \frac{dq_{2}}{%
dX_{3}} & = & 2q_{0}q_{1}%
\end{array}%
\right.
\end{equation}

These equations are coupled in pairs, so we can obtain an explicit solution
by just putting $q_{2}=q_{1}=0$. The system \eqref{sysq} reduces then to 
\begin{equation}  \label{sysqred}
\frac{dq_{0}}{dX_{3}} = -2q_{3}q_{0},\qquad \frac{dq_{3}}{dX_{3}} = n-q_{0}^{2}+q_{3}^{2}.
\end{equation}

An interesting result arises if we note that system \eqref{sysqred} can be
reduced to a Liouville type equation. In fact, making the substitution $%
q_{0}\equiv e^{2\alpha }=Y^{2}$, and writing $\tfrac{dY}{dX_{3}}\equiv
Y^{\prime }$, we obtain the ODE: $Y^{\prime \prime } =Y^{5}-nY$, 
that can be integrated to 
$[Y^{\prime\,2}-\frac{1}{3}Y^{6}+n Y^{2}]^{\prime }=0$,
from which we obtain the `energy equation', 
\begin{equation}
dY \left[C+\tfrac{1}{3}Y^{6}-nY^{2}\right]^{-\frac{1}{2}}=\pm dX_{3}.
\end{equation}
The above implicit relation is integrable but not invertible in general.
However, in the important particular case $C=0$, that is, the momentum map,
on-shell or surface constraint, the equation is fully solvable and
invertible. Thus, we obtain a non-commutative $N=2$ $BPS$ solution of the
form 
\begin{equation}  \label{quatwall}
q(X_{3})= (\sqrt{3n}) \operatorname{Sech}\left(\Phi\right) +%
\widehat{i^{3}} (i\sqrt{n}) \operatorname{Tanh}\left(\Phi)\right)
\end{equation}
where, $\Phi=2i\sqrt{n} X_3 -\beta_n$, and
 $\beta_n= 2\operatorname{Ln} \left( 3n \operatorname{e}^{i c_{0}\sqrt{n}}\right)$,
where $i$ is the imaginary unit of the field of complex numbers $\mathbb{C}$
and $c_0$ an arbitrary (complex) constant.

\begin{figure}[h!tb]\center%
\subfigure[ Hyperbolic (defect-like) case ($n<0$)]{\label{fig:s0s3xnNEG}
\includegraphics[{width=7cm}]{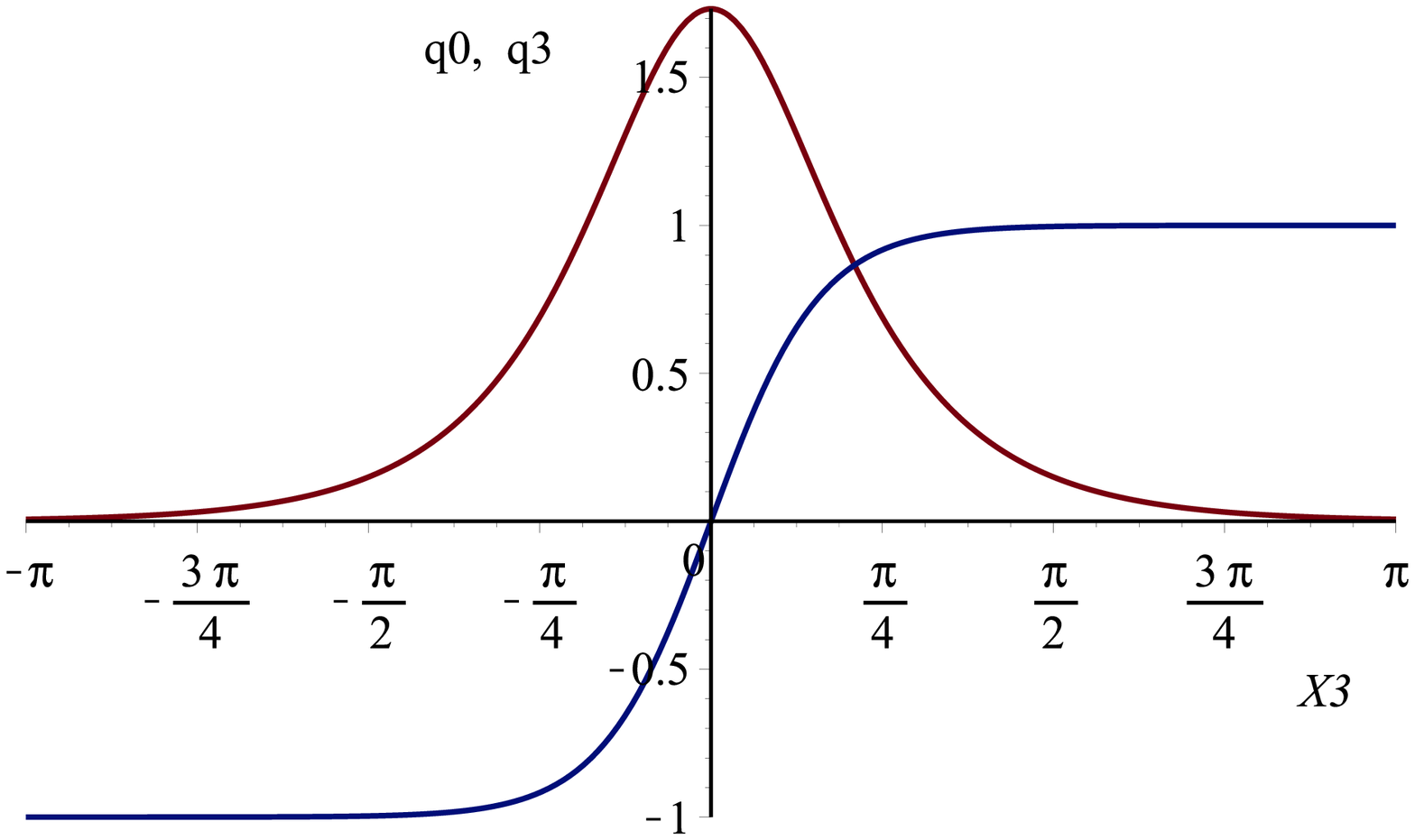}}\quad 
\subfigure[ Trigonometric case ($n>0$)]{\label{fig:s0s3xnPOS}
\includegraphics[{width=7cm}]{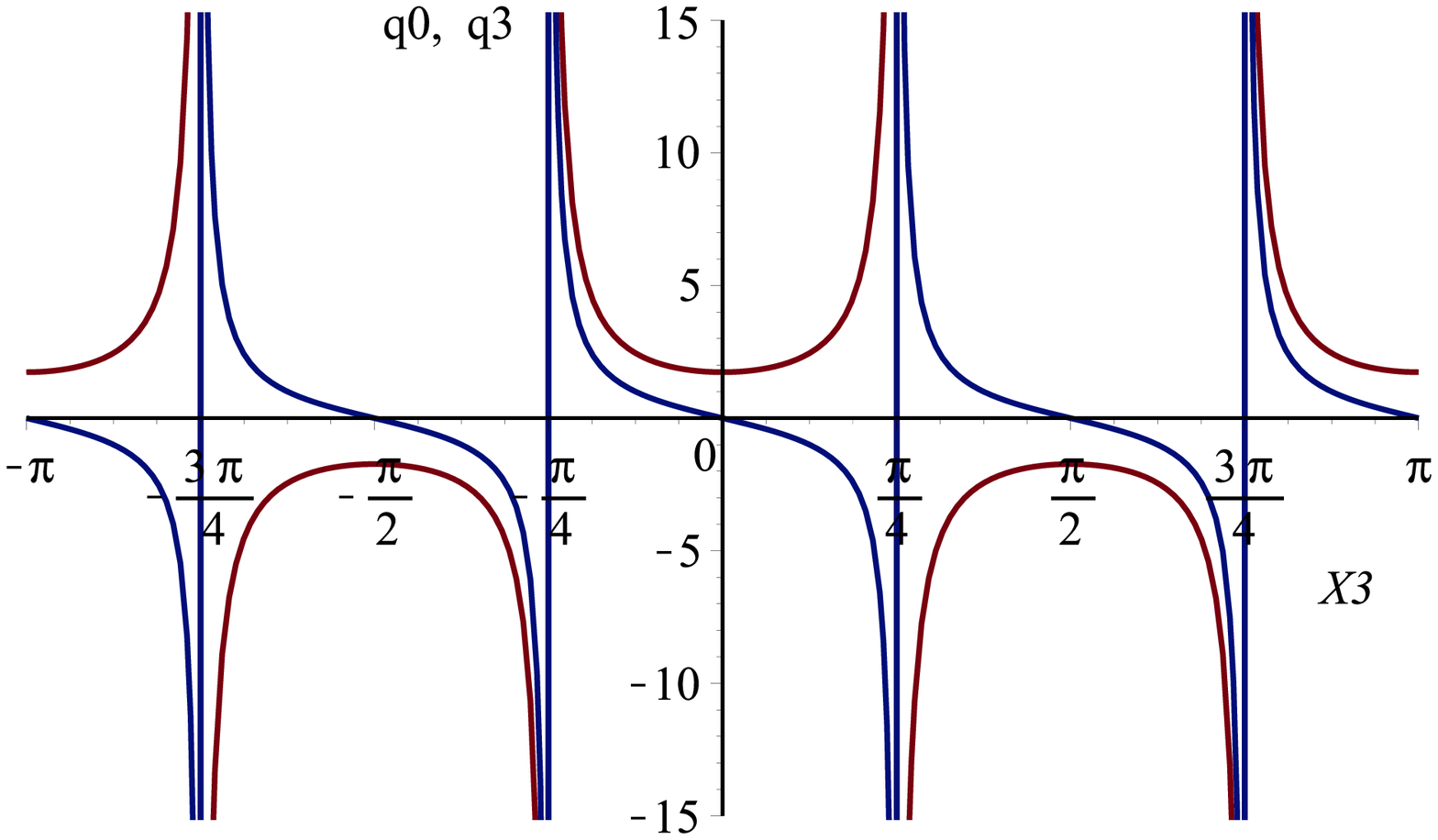}} \vspace{0.3cm}
\caption{Quaternionic solution. (a) $n=-1$, $c_0=\ln(3)$; (b) $n=1$, $c_0=i \ln(3)$.}
\label{fig:quatwall}
\end{figure}

It is worth noting here that, depending on the values of the constants $c_0$
and $n$ (and therefore $\beta_n$), solution \eqref{quatwall} can switch
between hyperbolic and trigonometric characters. In the context of
quaternionic $BPS$ structures developed in \cite{quat}, hyperbolic (negative 
$n$) case of solution \eqref{quatwall} can be interpreted as a domain wall
centered at $\left. X_{3}\right\vert _{0} = \beta_n/2$, showing the typical
defect behavior \footnote{Note that the pure imaginary profile of the scalar component of the solution
($q_0(X_3)$) is represented in Fig. \ref{fig:quatwall}, in the same graphic
than the real $q_3(X_3)$ component.} -- see figure \ref{fig:quatwall}.
Naturally, this defect-like behavior is lost in the trigonometric case.

As a result of the geometrical structure of the quaternionic solution \eqref{%
quatwall}, the potential \eqref{VN2} for the $N=2$ case of our $QWZ$ model
takes the form 
\begin{equation}  \label{VY}
V[q(X_3)]=2n^2 \operatorname{Sech}^2(\Phi) \left(4 \operatorname{Sech}^2(\Phi)-3 \right).
\end{equation}

\begin{figure}[h!tb]\center
\subfigure[Case $n<0$]{\label{fig:potq}\includegraphics[{width=5.5cm}]{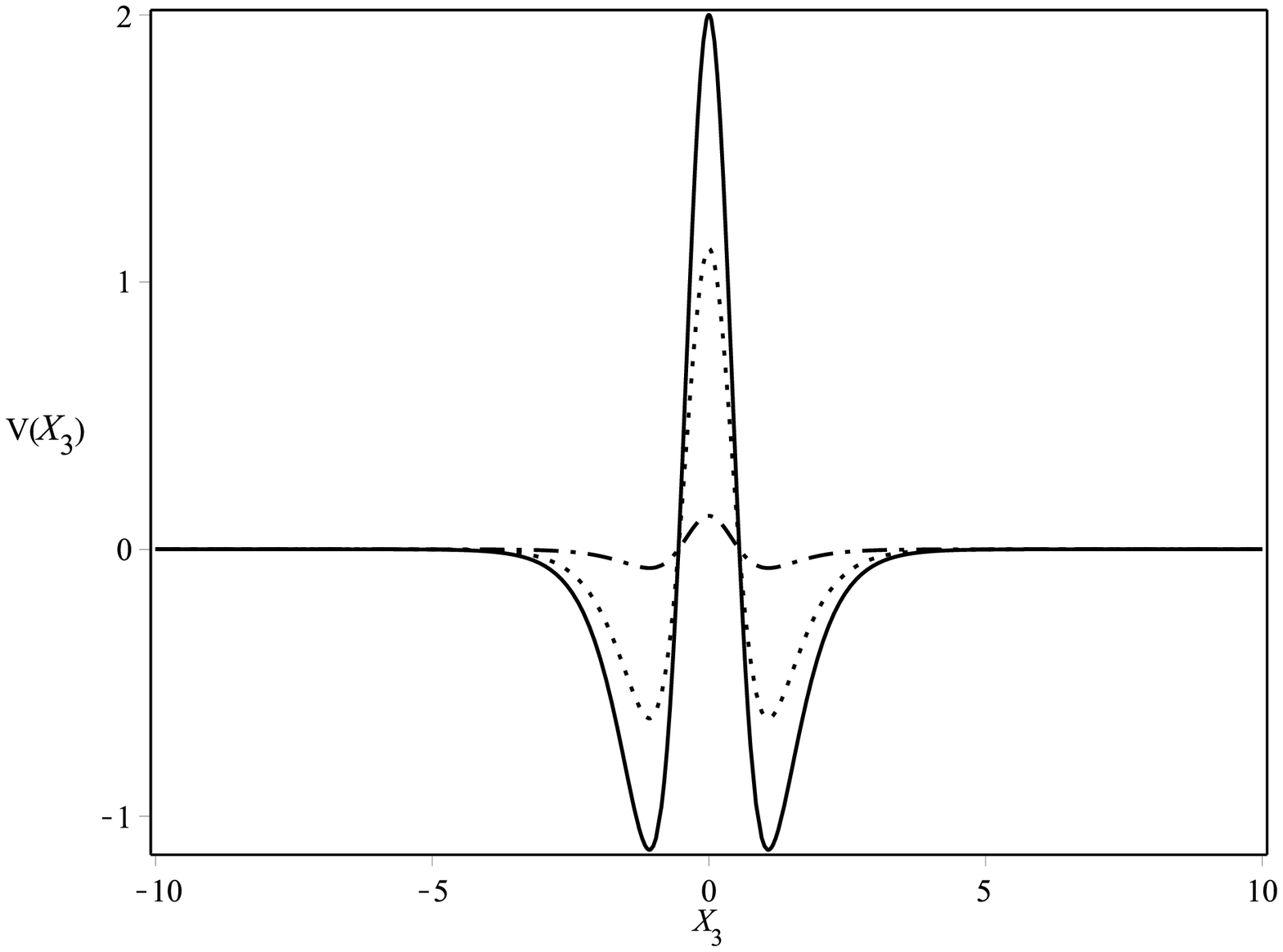}} %
\subfigure[Case $n>0$]{\label{fig:potn03}\includegraphics[{width=4.9cm}]{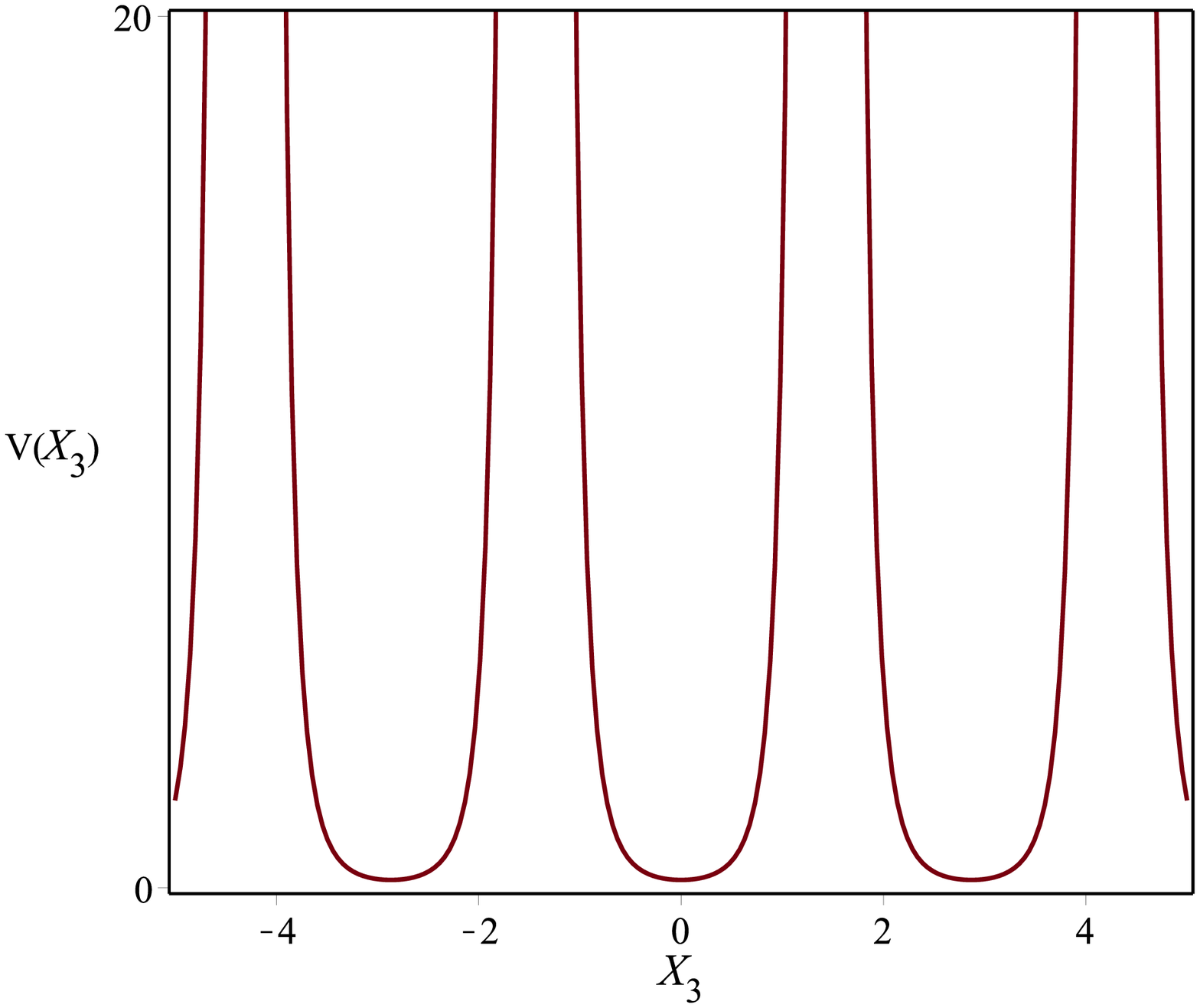}} %
\subfigure[Case $n>0$]{\label{fig:potn0325}\includegraphics[{width=5.2cm}]{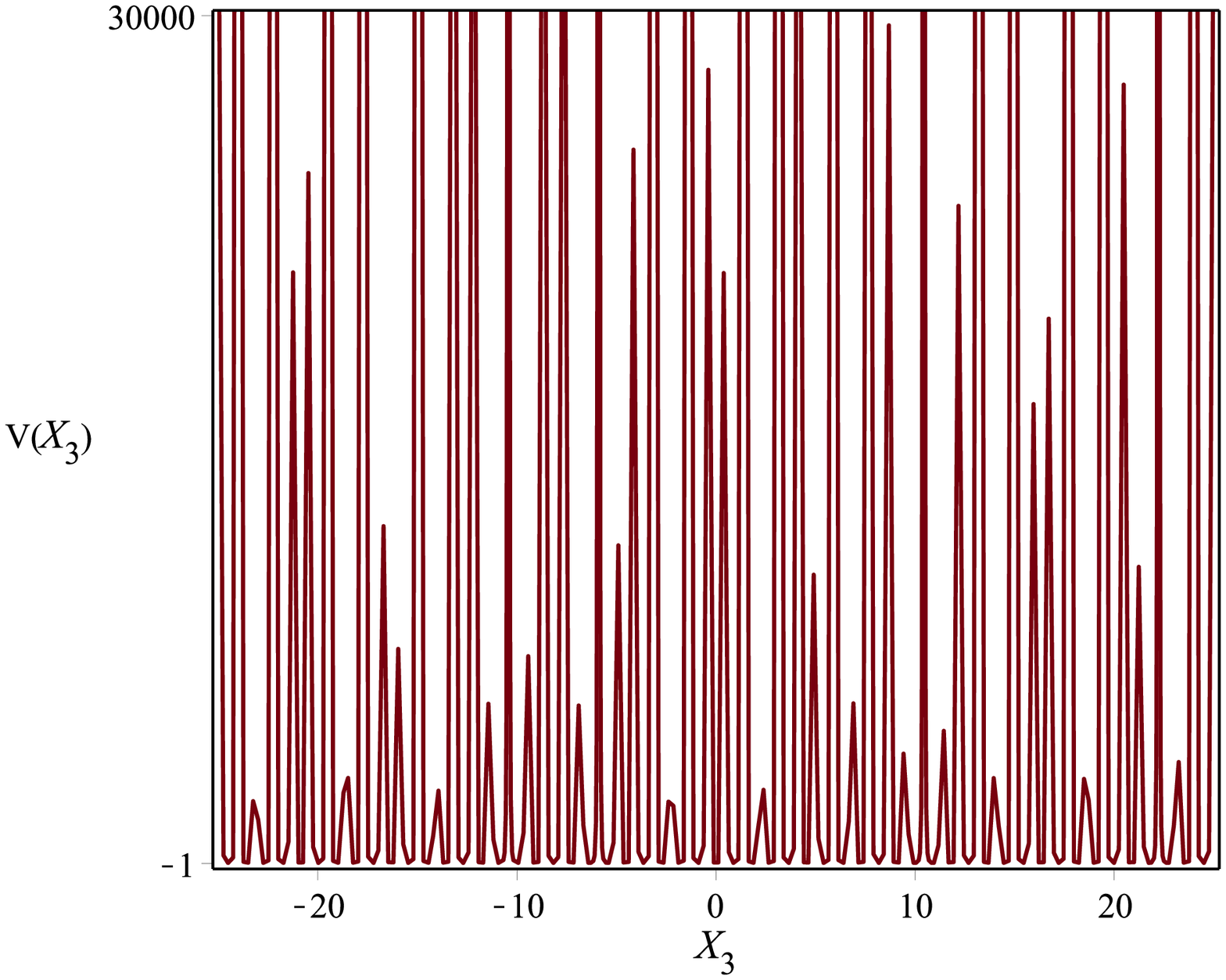}} \qquad
\caption{Spatial profile of quaternionic potential $V[q(x)]$. (a) `Inverted
volcano' potential ($n=-1, -0.75$ and $-0.25$). Figs. (b) and (c) show two
different scales of the pattern appearing in the $n>0$ (trigonometric) case.}\vspace{-2mm}
\label{fig:Vqx}
\end{figure}

For the sake of illustration, let us particularize the parameters $n$ and $%
c_0$ in order to get $\beta_n=0$. Then, the profile of the potential %
\eqref{VY} in terms of the base space variable $X_3$ take the different
forms depicted in Fig.\ref{fig:Vqx}. 
			
{\bf 4.3. Quaternionic model deformation.}
In complete analogy with the complex case considered in previous sections, 
our generalized quaternionic Lagrangian \eqref{GQL}, can be 
explicitly connected to a free Sigma model.

\smallskip{\it (a) Free quaternionic model.} As stated in \cite{quat}, the
connection (harmonic map) between the models can be reduced to the relation 
\vspace{-2mm}
\begin{equation}
V = [det(g_{ab})]^{-1}\equiv U^{-1},\vspace{-1.7mm}
\end{equation}
where $g_{ab}$ is the metric associated to the hyper-Kähler manifold (target
space) of the free model and $U$ is related to the line element as shown
below. Thus, the metric determinant in the $N=2$ case reads 
\begin{eqnarray}
\det\left( g_{ab}\right) &=&\left[\tfrac{1}{2}\left[n-(q_0^2-q_3^2)\right]^2+2 q_0^2 q_3^2\right]^{-1} \\
&=&\frac{\operatorname{Cosh}^2(\Phi)}{2n^2\left(1- 4%
\operatorname{Tanh}^2(\Phi)\right)}
\end{eqnarray}
where in the last equality we have made explicit the base space coordinate
via solution \eqref{quatwall} ($\Phi=2i\sqrt{n} X_3 -\beta_n$).

\smallskip{\it (b) Geometry of the wall: the line element.} In order to make
the corresponding geometrical analysis, let us introduce the obvious
transformation $\Theta=\operatorname{Tanh}\Phi$, which allows passing from
hyperbolic/trigonometric to polynomial expressions. Then solution %
\eqref{quatwall} and potential \eqref{VY} take the form 
\begin{eqnarray}  \label{VYTheta}
q(\Theta)&=&\widehat{i^{0}}\,(\sqrt{3n}) \sqrt{1-\Theta^2} + \widehat{i^{3}}%
\,(i\sqrt{n})\, \Theta \\
V[\Theta]&=&2n^{2}\left( 1-\Theta ^{2}\right) \left( 1-4\Theta
^{2}\right)=[det(g_{ab})]^{-1}
\end{eqnarray}

Now, taking into account the specific form of the line element in the hyper-K%
ähler and quaternionic cases we can write 
\begin{eqnarray}
ds^{2} &=&\! U^{-1}dq_{0}\otimes dq_{0} +U\, dq_{3}\otimes dq_{3} \\
&=&\! 12 n^{2}(1-\Theta^{2}) \Theta^{2}\! 
\left[U^{-1}\:\widehat{i^{0}}\otimes \widehat{i^{0}}
+\tfrac{1-\Theta^2}{3 \Theta^2} U\: 
\widehat{i^{3}}\otimes \widehat{i^{3}} \right] \nonumber
\end{eqnarray}
Therefore, we have 
$g_{00}=-24 n^{4} \Theta^{2} \left(1-\Theta ^{2}\right)^{2}\left(
1-4\Theta^{2}\right)$, and 
$g_{33}= 
{2\left(1-\Theta^{2}\right)}/{\left( 1-4\Theta ^{2}\right)}$.

\smallskip{\it (c) `Geometric map' and deformation.} As a final illustration, let us
propose the problem of finding a relation (deformation function) connecting
our hyper-Kähler wall solution \eqref{quatwall} to the solution obtained by
Arai, Nitta and Sakai (ANS) in \cite{nitta} --see also \cite{nittatalk}--, where
the geometry specified by metric determinant 
\begin{equation}
g_{\rm ANS}\equiv \left. \det \left( g_{ab}\right) \right\vert _{\rm ANS}=\tfrac{\mu
^{2}}{1-Q_{3}^{2}}  \label{gnitta}
\end{equation}%
is related to the solution 
\begin{equation}
Q_{3}(y)=\operatorname{Tanh}\left( \mu \,(y+y_{0})\right) ,  \label{qnitta}
\end{equation}
One way to compare \eqref{qnitta} to our wall solution 
\begin{equation}
q_3(X_{3})= (i\sqrt{n})\Theta = (i\sqrt{n})\operatorname{Tanh}\left(\Phi\right)\,,
\end{equation}
where $\Phi\equiv 2i\sqrt{n}X_{3}-\beta _{n}$,
is to write both expressions in the same coordinate basis. This is
accomplished by adjusting the coefficients taking $2i\sqrt{n}=\mu$ and $%
y_{0}=i\beta _{n}/(2\sqrt{n})$, and then identifying 
\begin{equation}  \label{identif}
X_3 \;\leftrightarrow\; y \qquad\text{and} \qquad Q_3 \;\leftrightarrow
\;\Theta= -i n^{-\frac{1}{2}} q_3,
\end{equation}%
Now the metric determinants explicit the distinct curvatures of the
solutions 
\begin{equation}
g_{ANS} =-\frac{4n}{1-\Theta^2}, \quad g=
\frac{1}{2n^2\left( 1-\Theta^2\right)\left(1-4\Theta^2\right)}.
\end{equation}
Then, 
\vspace{-2mm}\begin{equation}
F^\prime\!\left(\Theta\right)^2 = \frac{g(\Theta)}{g_{ANS}(\Theta)} 
= \frac{-1}{8 n^{3} \left(1-4\Theta^{2}\right)}
\end{equation}
from which we have 
\begin{equation}
F\left(\Theta\right) = -\frac{\sqrt {2}}{8{n}^{3/2}} \operatorname{Ln}\left[\sqrt{n}
\left( 2\Theta+\sqrt{4\,\Theta^2-1}\right)\right]+C_1.
\end{equation}

In order to get the function mapping one solution into the other,
that is, $f$ such that $\Theta =f(Q_{3})$, we use once again relation %
\eqref{geomfprime}. Then, under conditions \eqref{identif}, we can write the
derivative of the deformation function as 
\begin{eqnarray}
f^{\,\prime }\!\left( Q_{3}\right)^{2}&=&\frac{%
g_{ANS}(Q_{3}(X_{3}))}{g(f(Q_{3}))}\\
&=&-8n^{3}\,\frac{\left(
1-f(Q_{3})^{2}\right) \left( 1-4\,f(Q_{3})^{2}\right) }{1-Q_{3}^{2}}. \nonumber
\end{eqnarray}%
Integrating this expression we get, besides the trivial constant solutions $f(Q_{3})=\pm 1,\pm\frac{1}{2}$, 
\begin{equation}	
f(Q_{3})=i\sqrt{n}\operatorname{Sn}\!\left[ 
\sqrt{8n^3}\operatorname{Ln}\left(\sqrt{n}\,\Xi\,\right) +c_{1}\right],
\end{equation}%
where $\Xi\equiv Q_{3}+\sqrt{Q_{3}^{2}-1}$.
This is the function mapping the solutions of the two models ($q_{3}=i\sqrt{n}\Theta =i\sqrt{n}f(Q_{3})$) 
obtained by our geometric version of the deformation recipe. 

\smallskip{\bf 5. Fubini-Study space deformation. }

Let us now extrapolate the results obtained up to this point 
to a case not involving defect-like solutions, that is, in which we are not
worried about connecting the asymptotic values of the solutions of the
original and deformed models. In particular, in connection with our previous
results \cite{quat}, we analyze the Fubini-Study ($FS$) space case.

\vspace{1mm} One important (and usually disregarded) \emph{rôle} of the Fubini-Study
metric arises in the context of Quantum Mechanics. The Hilbert space
description of quantum theory is generically complex and, in order to
compute the transition probabilities, a (complex) scalar product is defined.
This scalar product determines an Euclidean geometry. However, another geometry
also emerges in this setting: the Fubini-Study geometry, which arises as follows.

As the superposition principle is assumed to be valid in quantum processes,
the theory must be linear, and two linear dependent vectors represent the
same physical state. Restricting to unit vector reduces the redundancy to
phase factors. Consequently, two curves on the unit vectors space --a curve
could be, for instance a piece of a solution of the Schrödinger equation--
differing only in a phase, describe the same set of physical states. In this
context, the `Fubini-Study length' corresponds to the minimal length a curve
of states can assume. The requirement of the minimal length on the set of
vector states induces a geometry, represented by the `Fubini-Study metric' 
\cite{uhlmann}. Imposing the condition that the Euclidean and the
Fubini-Study lengths coincide piecewise in the curve (parallel transport
condition) defines the geometric (or Berry) phase, which corresponds to the
difference between an arbitrary solution states (closed) curve and the one
obtained imposing the minimal length condition. More precisely, its initial
and final points will differ in a phase factor: the `geometric phase'.

\smallskip{\bf 5.1 $FS$ space from a complex field model. }
The spaces described above are realized by Kähler Manifolds
(complex manifold $M$ with a Hermitian metric $h$ and a fundamental closed
2-form $\omega$), with group structure $\mathbb{CP}^{n} = S^{2n+1}/ S^{1}.$
The Hermitian metric components are given by 
\begin{equation}  \label{Hmetric}
h_{i\overline{j}}\equiv h\left( \partial _{i}\overline{\partial _{j}}\right)
=\frac{( 1+\left\vert z\right\vert ^{2}) \delta _{i\,\overline{j}%
}-z_{i}\overline{z_{j}}}{( 1+\left\vert z\right\vert ^{2}) ^{2}}.
\end{equation}%
Then the Kähler potential ($h_{i\overline{j}}=\partial_i\bar\partial_j K(z,\bar z)$) reads
\begin{equation}  \label{K}
K(z, \bar z)=\ln (1+\delta _{i\,\overline{j}}z^{i}\overline{z^{j}}).
\end{equation}

Consider now the Lagrangian density 
\begin{equation}  \label{L1plusphi}
\tilde{\mathcal{L}_{\mathcal{K}}}
=\frac{1}{2}\partial _{\mu }\tilde{\varphi}%
\overline{\partial ^{\mu }\tilde{\varphi}}-\left(1+\left\vert \tilde{\varphi}\right\vert ^{2}\right) ^{2}
\end{equation}%
From the form of the potential term and relation \eqref{eq:freecomplexb}, it is straightforward to conclude that
the target space of the free version of the model can be related to a Fubini-Study space with a Khäler potential as in \eqref{K}.

Let us first analyze the model \eqref{L1plusphi}. 
Putting the field in its polar representation, $\tilde\varphi \equiv r e^{i\alpha }$, 
the Lagrangian takes the form
\begin{equation}
\tilde{\mathcal{L}_{\mathcal{K}}}=\frac{1}{2}\left[ \left( \partial _{\mu
}r\partial ^{\mu }r\right) +\left( \partial _{\mu }\alpha \,\partial ^{\mu
}\alpha \right) r^{2}\right] -\left( 1+r^{2}\right) ^{2}.
\end{equation}%
The corresponding Euler-Lagrange equations are 
\begin{eqnarray}
\partial ^{\mu }\left( r^{2} \partial _{\mu}\alpha \right) &=&0\qquad  \label{eqa}
\\
\partial^{\mu}\partial_{\mu} r-\left( \partial _{\mu }\alpha\,\partial^{\mu}\alpha\right)r
+4r\left( 1+r^{2}\right) &=&0  \label{eqr}
\end{eqnarray}%
Equation \eqref{eqa} above expresses the existence of a conserved quantity
\begin{equation}
L_{\mu }\equiv r^{2} \partial _{\mu }\alpha.
\end{equation}
Substituting $L_\mu$ in 
\eqref{eqr}, we have 
\begin{equation}
\partial ^{\mu }\partial _{\mu }r-\frac{L_{\mu }L^{\mu }}{r^{3}}+4r\left(1+r^{2}\right) =0
\end{equation}

\begin{figure}[h!tb] 
\includegraphics[width=4cm]{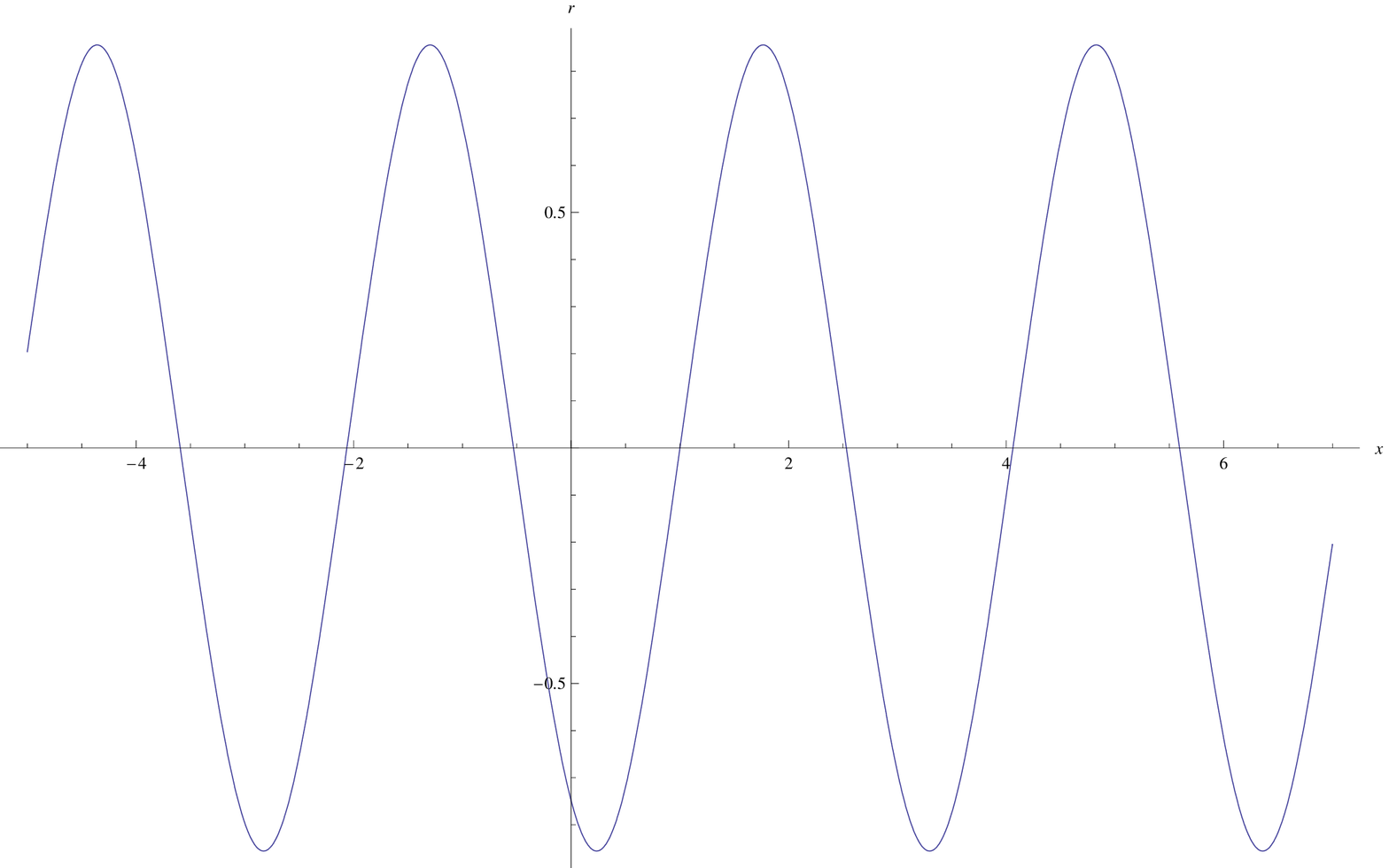}
\caption{$r(x)$ solution with $\alpha=const$ (`momentum map').}
\label{figrL0}
\end{figure}

The simplest case we can consider to obtain a solution for this
equation is the `momentum map' ($L_{\mu }=0$ or $\alpha=const.$). In that case, after covariant
elimination, we arrive to the solution 
\begin{equation}
r=-\operatorname{Re}\{ i\; c_{-} \operatorname{sn}\left(i wx, c_{+}/c_{-}\right)\},
\end{equation}%
where $w x = w_{\mu }x^{\mu }$, $c_\pm=1\pm\sqrt{[1+2c]}$ and $c$ and $w_{\mu }$ are scalar and vector
constants, respectively.
The momentum map $L_{\mu }=0$ introduces a degeneration of the solution $%
\tilde{\varphi}=\overline{\tilde{\varphi}}=r $, which results real,
continuous and periodical, showing the typical behavior of the $\operatorname{sn}$ -- see Fig. \ref{figrL0}.

\begin{figure}[h!tb]
\label{fig:Lneq0} \center
\subfigure[]{\label{fig:rLneq0}\includegraphics[{width=3.5cm}]{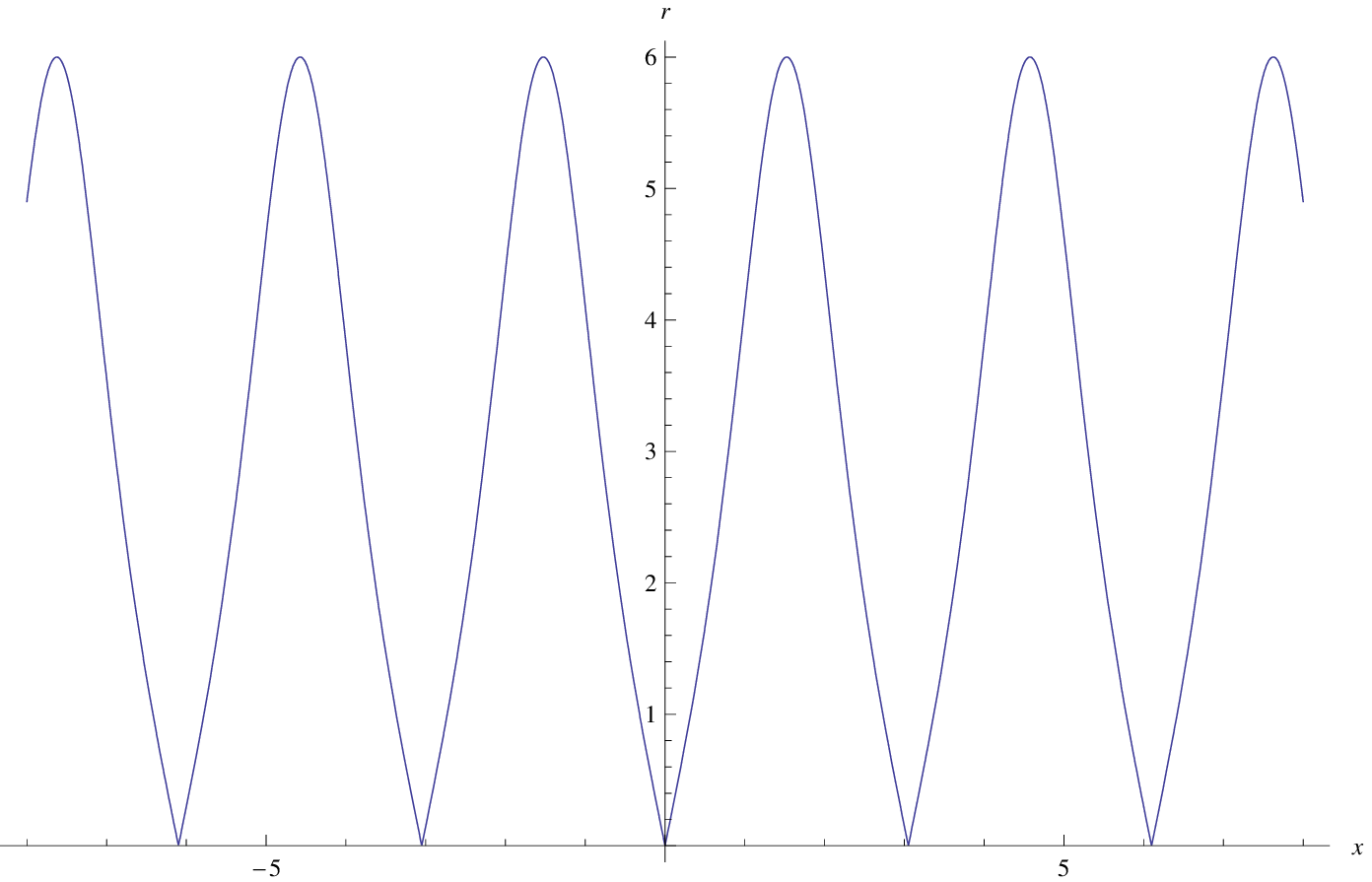}}\quad %
\subfigure[]{\label{fig:aLneq0}\includegraphics[{width=4cm}]{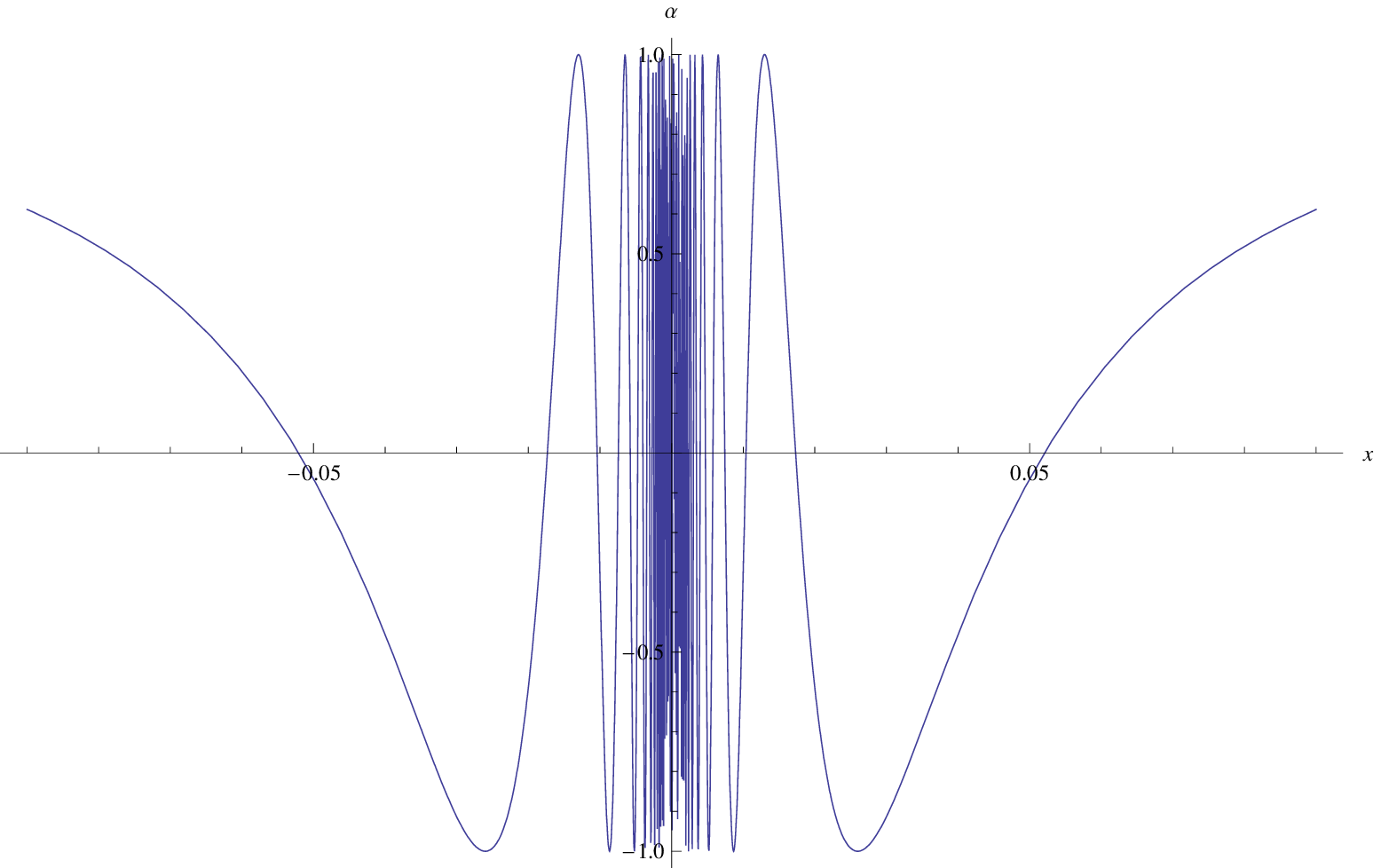}}%
\caption{ $L_\protect\mu\neq0$ solution. (a) Norm $r(x)$; (b) Highly
oscilating phase $\protect\alpha(x)$.}
\end{figure}

In the general case ($L_\mu\neq0$), the norm of the scalar field takes the form 
\begin{equation}
r=\!-\operatorname{Re}\left\{\operatorname{sn}\left(i\lambda wx,\tfrac{\lambda_{-}}{\lambda_{+}}\right) 
\left[\lambda_{+}\!-\operatorname{sn}^2\left( i\lambda wx,\tfrac{\lambda_{-}}{\lambda_{+}}\right)\right]^{\frac{1}{2}} \right\}
\end{equation}
where $\lambda_{\pm}\equiv 1 \pm L_{\mu }L^{\mu }/4$ and $\lambda=\sqrt{%
\lambda_+/2}$. This solution, depicted in Fig. \ref{fig:rLneq0}, is periodic
with compact support on the spacetime coordinates (we can, for instance, identify 
$x$ with the radial coordinate of a cylindrical spacetime).

The phase of the general solution can also be explicitly determined however,
the expression is cumbersome and we will just mention here that it presents
a highly oscillating behavior near the origin of coordinates, and stabilizes
quickly to $\alpha\rightarrow \sqrt{L_{\mu }L^{\mu }}$ as moving away 
(See Fig. \ref{fig:aLneq0}). 

{\bf 5.2. Fubini-Study space from a flat space.}
 We have shown an explicit complex field solution in connection with a
Fubini-Study space. Consider now a one dimensional ($n=1$) flat complex
space, described by a flat metric $g_{flat}$. From relation \eqref{eq:freecomplexb}, 
the corresponding Lagrangian density with a potential
reads \footnote{As before, we are considering the simplest $\mathcal{C}_{0}\!=\!1$ case.}
\begin{equation}  \label{Lk}
\mathcal{L}_{\mathcal{K}}=\frac{1}{2}\partial_{\mu}\varphi\overline{%
\partial^{\mu}\varphi}-V(\varphi,\overline{\varphi})=\frac{1}{2}%
\partial_{\mu}\varphi\overline{\partial^{\mu }\varphi}- \overset{%
g_{flat}^{-1}}{\overbrace{m^{2}}},
\end{equation}
That is, $V(\varphi, \bar\varphi)=m^2=const.$ 
Then, using the precedent results, we can explicitly link model \eqref{Lk} to a
Fubini-Study space, which corresponds to the target space of the free
version of the Lagrangian \eqref{L1plusphi}. That is, taking 
\begin{equation}\label{eq:}
g_{flat}^{-1}\leftrightarrow V=m^{2} \quad\text{and} \quad
\tilde{g}^{-1}_\text{FS} \leftrightarrow \tilde{V}=(1+\left\vert \tilde{\varphi}\right\vert^{2})^{2}
\end{equation}
Now, as stated in \eqref{eq:fcplx}, the function $f$ connecting $\varphi$ and 
$\tilde\varphi$ satisfies
\begin{equation}
|f^{\,\prime}(\tilde{\varphi})|^{2}= \left\vert \frac{d\varphi}{d\tilde{\varphi}}\right\vert^{2} 
=\frac{\tilde{g}_\text{\,FS}}{g_{flat}} =\frac{m^{2}}{(1+\left\vert \tilde{\varphi}\right\vert ^{2})^{2}},
\end{equation}
that is, $d\varphi =m\, d\tilde{\varphi}\, /(1+ i \tilde{\varphi})^{2}$, 
which directly leads to relation 
\begin{equation}\label{fFS}
\varphi = f(\tilde\varphi)= \frac{m}{\tilde\varphi-i}=\frac{m}{(1+\vert \tilde{\varphi}\vert ^{2})} (\overline{\tilde\varphi}+i)
\end{equation}

Relation \eqref{fFS} above can be easily check against the results of previous
sections  by reconstructing the deformed potential applying \eqref{complexdefV} to get
\begin{equation}
\frac{V(f(\tilde\varphi),
\overline{f(\tilde\varphi)})}{\left\vert f^{\prime}(\tilde\varphi)\right\vert ^{2}}=
\frac{{m^{2}}}{{m^{2}}/{(1+\vert \tilde{\varphi}\vert^{2})^{2}}}
\!\!=(1+\left\vert \tilde{\varphi}\right\vert ^{2})^{2}
=\tilde{V}(\tilde{\varphi},\bar{\tilde{\varphi}})\notag	
\end{equation}
Thus, at least for the present case, the procedure can be
applied in a purely geometric context, \emph{i.e.} when there are no defect
solutions associated.

\medskip{\bf 6. Concluding Remarks.}

In the present article we have considered a technique for generating
solutions, known as deformation method. Differently from all the previous
works on this subject, here we analyze the procedure under a geometrical
point of view, in order to get a better understanding of the meaning and
limitations of the method. Thus, the main idea of our analysis was developed
working with the metric fields of the equivalent free models, connecting
different solutions and also generating new ones.

Two main results are to be remarked: First, we have shown through this paper
that the `deformation procedure' is nothing but the simplest
geometrical map between free Lagrangians in curved spacetimes,
equivalent to the respective flat space Lagrangians with a potential.
Second, as put in evidence by the geometrical treatment, we conclude that
the deformation method is unavoidable limited to systems with higher
symmetries and can be consistently implemented only in the case of
single-field models. This is due, mainly, to the very restrictive constraint
imposed by preservation of $BPS$ conditions (first order equations).

As a way to delimitate the widest applicability of the procedure, systems with
multiple field components were considered. Namely, real ($\mathbb{R}$),
complex ($\mathbb{C}$) and also quaternionic ($\mathbb{H}$) one-dimensional
field models were analyzed. All cases where illustrated by constructing
explicit maps between different associated geometries.

The present work is part of a larger investigation, centering efforts in
finding and studying new quaternionic $BPS$ solutions. In that direction, 
a novel quaternionic domain wall solution was obtained here by directly
performing a geometrical map. 

In constructing domain wall solutions, the method shows highly suitable due to
the kind of geometries related in the transformations. 
However, our geometrical approach put in clear evidence the strong limitations 
of the deformation procedure as a tool for generating solutions, as it is the 
simplest harmonic map between the metrics
(of the geometrical Lagrangians associated to the mapped dynamical models).
Thus, in more involved geometries-field models, the use of pure geometrical
approach to the procedure here presented is mandatory in order to preserve $BPS$ conditions, 
SUSY structure or gauge transformations, as was clearly illustrated by the mapping of the 
Fubini-Study geometry in Section 5. 

Note also that we have avoided the problem of preservation of the Noether charges 
({\em e.g.} the $U(1)$-charge in the complex case) arising when proposing the simplest possible 
mappings between the interacting and the free models,
by working all the examples in the momentum map.

Differently from the `standard' mapping of references \cite{defmet, otherdef}, which
does not assures the preservation of such symmetries, the geometrical
approach could, in principle, be extended to a $N=2$ gauged supergravity theory in four
dimensions. This should be possible given that a hyper-Kähler manifold $\mathcal{M}_{H}$ 
is a quaternionic space in which, as shown in \cite{quat}, and also in Sec. 4.1,  
stable $BPS$ solutions can be consistently constructed.

Another interesting potential feature of the geometrical treatment of the mappings 
shown through the present work, is the possibility of establishing a parallel between effective models 
with modified kinetic term (`k-fields'), of interest to the dark energy-matter problem, 
and models described by Lagrangians of the standard form. 
These idea is under exploration and will be presented elsewhere.

\medskip{\bf 7. Aknowledgments.}
DJCL would like to thank CONICET (Argentina) and JINR (Russia), and VIA thanks PROCAD/CAPES (Brasil).


\end{document}